\documentclass[a4paper,11pt]{article}
\usepackage{pos}
\usepackage{xcolor}

\renewcommand{\Im}{\mathrm{Im}\,}

\title{Real-time dynamics from convex geometry}

\author*{Scott Lawrence}

\affiliation{Los Alamos National Laboratory Theoretical Division T-2,\\Los Alamos, NM 87545, USA}

\emailAdd{srlawrence@lanl.gov}

\abstract{A quantum-mechanical system comes naturally equipped with a convex space: each (Hermitian) operator has a (real) expectation value, and the expectation value of the square any Hermitian operator must be non-negative. This space is of exponential (e.g.~in volume) dimension, but low-dimensional projections can be efficiently explored by standard algorithms. Such approaches have been used to precisely constrain critical exponents of conformal field theories ("conformal bootstrap") and, more recently, to constrain the ground state physics of various quantum-mechanical systems, including lattice field theories. In this talk we discuss related approaches to systematically constraining the real-time dynamics of quantum systems, which are otherwise obstructed from study by sign problems and the ill-posed nature of analytic continuation.}

\FullConference{The 41st International Symposium on Lattice Field Theory (LATTICE2024)\\
 28 July - 3 August 2024\\
Liverpool, UK\\}

\begin{document}
\maketitle

\section{Introduction}

Lattice Monte Carlo calculations, performed in imaginary time, provide
estimates of the Euclidean correlator $G^{(E)}(\tau)$. These estimates are available at a finite number of time-separations, determined by the lattice spacing and the circumference of the thermal circle: $\tau_n = n \frac{\beta}{N_\tau}$ for $n \in \mathbb Z$. The exact Euclidean correlators, including data at all $\tau \in [0,\beta)$, generally uniquely determine all properties of the quantum field theory~\cite{Osterwalder:1973dx}. This fact does not imply that those properties are easily determined from the Euclidean correlators, and in particular says nothing about the task of determining various properties from the restricted, noisy access given by lattice Monte Carlo.

As a motivating example, consider linear response. A system is prepared in thermal equilibrium (or the ground state) of a Hamiltonian $H_0$, and thereafter evolved under a time-dependent Hamiltonian:
\begin{equation}
	H(t) = H_0 + \epsilon \delta(t) \mathcal O\text.
\end{equation}
Due to the perturbation $\epsilon$, the expectation value $\langle \mathcal O(t)\rangle$ has nontrivial time-dependence. In the limit of small perturbations this dependence is entirely captured by the two-point correlator:
\begin{equation}
	\langle \mathcal O(t) \rangle = \langle \mathcal O(0)\rangle_0 + 2 \epsilon G(t) + O(\epsilon^2)
		\quad\text{ where }\quad G(t) = \Im \langle \mathcal O(t) \mathcal O(0)\rangle_0
\end{equation}
Above, $\langle \cdot\rangle_0$ denotes expectation values with respect to the unperturbed ($\epsilon = 0$) system.

This talk is thus motivated by the task of determining the \emph{real-time} correlator $G(t)$. This is the analytic continuation of $G^{(E)}(\tau) = G(i\tau)$, and therefore this is a talk about a problem of analytic continuation---albeit, with extra structure provided by the axioms of quantum mechanics. The analytic continuation is uniquely determined from $G^{(E)}(\tau)$, but we do not have the full correlator. We are constrained in two separate ways: we have data only at a finite set of separations $\{\tau_i\}$, and the data are noisy.

Our problem of analytic continuation, then, is \emph{ill-posed}: there is not one function $G(t)$, but many which are consistent with the data available. This does not imply that no information about the real-time correlator is available. For example, it is straightforward to show (by writing each correlator as an explicit sum over states) that at any time $t$ we have the inequality $|G(t)| \le |G^{(E)}(0)|$. Tighter bounds may be derived with more effort. This talk concerns a systematic method of establishing bounds on the real-time correlator, given finite Euclidean data. Furthermore these bounds are the tightest possible from the given Euclidean data, assuming no knowledge besides quantum-mechanical axioms.

To organize the computation of these bounds, it is helpful to rewrite both the real-time and Euclidean correlators as integrals of the spectral density $\rho(\omega)$. Positivity of the norm on the Hilbert space ($\langle \mathcal O^\dagger \mathcal O \rangle \ge 0$ for any operator $\mathcal O$ and in any state) implies $\rho(\omega) \ge 0$ for all frequencies. Requiring that the spectral density be consistent with the Euclidean data further defines a convex space of spectral density functions. We may then use techniques of convex optimization to establish lower and upper bounds on any desired integral of the spectral density, including smeared real-time correlators. This final technical step, the solution of a convex optimization problem, has served a similarly key role in the conformal bootstrap~\cite{Kos:2016ysd,Poland:2018epd}, deriving constraints on various quantum-mechanical systems~\cite{Han:2020bkb,Berenstein:2021dyf}, constraints on lattice field theories~\cite{Lawrence:2021msm,Lawrence:2022vsb}, and most recently constraining the real-time dynamics of quantum-mechanical systems in the absence of lattice data~\cite{Lawrence:2024mnj}.

An alternative---and more common---approach to ill-posed analytic continuation problems is to select a regulator. In maximum entropy methods~\cite{Asakawa:2000tr}, the regulator is a decision to seek the spectral density that maximizes some entropy subject to the given Euclidean constraints. In cases where an explicit fit is performed~\cite{Tripolt:2018xeo,Bailas:2020qmv}, the regulator is implicit in the choice of functional form to fit to (which may be a neural network~\cite{Chen:2021giw,Wang:2021jou,fournier2020artificial,Kades:2019wtd,Horak:2021syv}). In all such methods, too numerous to properly review here, the choice of regulator may introduce a systematic bias into the results. (This same fact is also a great advantage to regulator-based methods: a well-chosen regulator may encode reasonable physical priors.)

This talk will necessarily be only a brief overview of the use of convex optimization to perform inferences from lattice data. The full details of the computations presented in this talk, and some extensions, are given in~\cite{Lawrence:2024hjm}.

\section{Geometry of the spectral density}
The Euclidean correlator (to which we have limited access via lattice Monte Carlo) and the real-time correlator are connected by analytic continuation, or alternatively, through the spectral density:
\begin{equation}
	G(t) = - \int d\omega\, \rho(\omega) \sin\omega t
\quad\text{ and }\quad
	G^{(E)}(\tau) = \int d\omega\, \rho(\omega) \frac{\cosh \omega \left(\frac \beta 2 - \tau\right)}{\sinh \frac{\beta\omega}2}\text.
\end{equation}
This relation allows us to rephrase our ``analytic continuation'' problem as an inversion problem: given $G^{(E)}(\tau)$, determine some properties of $\rho$. In this language, the constraint of Hilbert space positivity is easily encoded. It is equivalent to the statement $\rho(\omega) \ge 0$.

To make the structure of this problem clear, let us generalize. We are provided with noisy estimates of $N$ distinct integrals of $\rho$, specified by a set of kernels $\{K_i(\omega)\}$. These estimates have means $C_i$ defined by
\begin{equation}
	C_i \approx \int_0^\infty K_i(\omega) \rho(\omega) d\omega
	\text.
\end{equation}
The covariance matrix (which may in practice be determined by the statistical bootstrap) is denoted $\Sigma$. This captures the statistical errors on $C_i$. From this data we wish to determine an integral of $\rho$ defined by the integration kernel $\mathcal K$: $\int_0^\infty \rho(\omega) \mathcal K(\omega) d\omega$.

The space of possible integrals $C$ is (some subspace of) $\mathbb R^N$. 
The estimates $C_i$ and the covariance matrix $\Sigma$ together define an ellipsoid in $\mathbb R^N$: the set of all points $x \in \mathbb R^N$ such that a certain statistic $F$ obeys $F(x) \le F_{\mathrm{max}}$:
\begin{equation}\label{eq:statistic}
	F(x) = x^T \Sigma^{-1} x\text.
\end{equation}
The threshold $F_{\mathrm{max}}$ may be chosen to define a desired confidence interval---all plots in this talk are based on a 99\% confidence interval. This is accomplished by resampling the Monte Carlo data.

Now, define a functional $v[\rho]$ that takes a spectral density function and returns the point in $\mathbb R^N$ corresponding to the integrals being estimated:
\begin{equation}
	v_i[\rho] = \int_0^\infty \rho(\omega) K_i(\omega) d\omega\text.
\end{equation}
By means of this function, the above condition on points $x$ can be reframed directly as a condition on spectral density functions $\rho(\omega)$:
\begin{equation}
	F[\rho] = v[\rho]^T \Sigma^{-1} v[\rho] \le F_{\mathrm{max}}\text.
\end{equation}
Spectral density functions that obey this condition may be said to be consistent with the provided Euclidean data.

Putting all this together, we can write the task of finding the smallest possible value of $\int \mathcal K \rho$ in the form of a convex optimization problem:
\begin{equation}\label{eq:spectral-primal}
	\begin{split}
		\text{minimize }& \int_0^\infty d\omega\, \mathcal K(\omega) \rho(\omega)\\
		\text{subject to }& \rho(\omega) \ge 0 \\
		\text{and }& F[\rho] \le F_{\mathrm{max}}
		\text.
	\end{split}
\end{equation}
The solution to this convex program provides the minimum value of $\int \mathcal K \rho$ consistent with both positivity and the provided Euclidean data. A maximum value is obtained by re-solving with $\mathcal K \rightarrow -\mathcal K$.

At this point a serious problem arises. Although convex optimization is computationally tractable---polynomial in the sizes of all inputs---we have here an optimization problem operating on an infinite-dimensional space of functions. We cannot hope to solve this convex program on a computer of finite size. We might discretize $\omega$, effectively introducing some ansatz for $\rho$. This will artificially tighten the ``bounds'' obtained by solving the convex program: the true bounds are only recovered after an expensive limiting procedure.

Instead, we will pass to the \emph{Lagrange dual} of the above optimization problem (termed the \emph{primal}). The primal problem has infinitely many variables; we will see that the dual has finitely many, and therefore may be solved in finite time. First define a Lagrange functional as follows:
\begin{equation}
	L[\rho(\omega);\lambda(\omega),\mu] =
		\int \mathcal K(\omega) \rho(\omega) d\omega
		- \int \lambda(\omega) \rho(\omega) d\omega - \mu (1 - v[\rho]^T \Sigma^{-1} v[\rho])
	\text.
\end{equation}
Note that the Lagrange multipliers $\lambda$ and $\mu$ are different sorts of objects: $\mu$ is simply a real number, while $\lambda$ is a function (in the dual space to the space of spectral density functions $\rho$). The convex optimization problem (\ref{eq:spectral-primal}) above can now be written simply as
\begin{equation}
	p^* = \min_{\rho(\omega)} \max_{\mu,\lambda(\omega) \ge 0} L[\rho;\lambda,\mu].
\end{equation}
The optimal value $p^*$ is often called the \emph{primal optimum}, to contrast with the \emph{dual optimum} obtained by swapping the extremizations:
\begin{equation}\label{eq:dual-optimum}
	d^* = \max_{\mu,\lambda(\omega) \ge 0} \min_{\rho(\omega)} L[\rho;\lambda,\mu]
	\text.
\end{equation}
It is straightforward to prove that $d^* \le p^*$. Less obvious is that $d^* = p^*$: sufficient conditions are given by Slater~\cite{RePEc:cwl:cwldpp:80}, and are satisfied by (\ref{eq:spectral-primal}). Therefore, by solving the dual problem we obtain also the primal optimum, and the desired lower bound on $\int \mathcal K \rho$.

We now proceed to obtain the dual optimization problem by analytically performing the inner minimization in (\ref{eq:dual-optimum}). At first glance, it is not clear that this is an improvement over the primal: after all, we will have a maximization over another functional space, this time of $\lambda(\omega)$. But in fact, the inner minimization (over $\rho(\omega)$) turns out to impose affine constraints on $\lambda(\omega)$. In particular $\lambda$ must be of the form
\begin{equation}
	\lambda(\omega) = \mathcal K(\omega) - \sum_i \ell_i K_i(\omega)
	\text,
\end{equation}
for some real coefficients $\ell_i$.
For any $\lambda(\omega)$ that is \emph{not} of this form, the inner minimization returns $-\infty$ (making that point irrelevant to the outer maximization).

This fact is key to the usefulness of the Lagrange dual. Only a finite-dimensional space of functions $\lambda(\omega)$, parameterized by $\ell_i$, are ``permitted'', in the sense of being candidates for producing the dual optimum. We may therefore analytically evaluate the inner minimization on this restricted space to finally obtain the dual program (see~\cite{Lawrence:2024hjm} for a complete derivation):
\begin{equation}
\label{eq:spectral-dual}
	\begin{split}
		\text{maximize }&\ell^T C - \frac{F_{\mathrm{max}}}{4\mu} \ell^T M^{-1} \ell - \mu\\
		\text{subject to }& \mathcal K(\omega) - \sum_i \ell_i K_i(\omega) \ge 0\\
		\text{and }&\mu \ge 0
	\text.
	\end{split}
\end{equation}
The same technique for converting an infinite-dimensional optimization problem to a finite-dimensional one is at work in the so-called ``real-time bootstrap''~\cite{Lawrence:2024mnj}.

Before moving from this general case back to the specifics of the real-time correlator, note that with $N$ pieces of Euclidean data (e.g.~$N$ time-slices), the convex optimization problem to be solved has only $N+1$ dimensions. The time taken to solve a convex program of this form in practice is typically around $\sim N^3$. Thus even on unusually large lattice, computing a lower bound in this way is cheap---certainly relative to the lattice calculation itself!

\section{Extracting the real-time correlator}
We now return to the specific problem at hand: extracting the real-time correlator given noisy estimates of the Euclidean correlator. Before proceeding we must set expectations. The real-time correlator itself cannot be meaningfully bounded. Specifically, the only bound on $G(t)$ which is possible is the bound $|G(t)| \le |G^{(E)}(0)|$. This is because the high-energy behavior of $\rho(\omega)$ is not visible in the Euclidean correlator at any non-zero $\tau$.

Therefore if we hope to obtain interesting bounds, we must ask a question which is not sensitive to the behavior of the theory at high energies. This is naturally accomplished by defining a smeared real-time correlator as follows:
\begin{equation}
	\tilde G_{\sigma}(t) \equiv \frac{1}{\sqrt{2\pi}\sigma} \int_{-\infty}^\infty dt'\,G(t)e^{-(t'-t)^2 / (2 \sigma^2)}\text.
\end{equation}
This smearing strongly damps the influence of frequencies $\omega$ for which $\omega\sigma \gg 1$.

To relate to the general form of the inversion problem detailed previously, we define the integration kernels $K_i$ and $\mathcal K$ to be
\begin{equation}
K_i(\omega) \equiv \frac{\cosh \omega \left(\frac \beta 2 - \tau_i\right)}{\sinh \frac{\beta \omega}{2}}
	\quad\text{ and }\quad
\mathcal K^t_\sigma(\omega) \equiv -
		e^{-\frac{\sigma^2 \omega^2}{2}}
		\sin \omega t
\text.
\end{equation}
For each choice of $(t,\sigma)$, these kernels define a convex optimization problem via~(\ref{eq:spectral-dual}), which bounds the real-time correlator at $t$, smeared with a Gaussian width of $\sigma$.

We now proceed by solving the convex optimization problem~(\ref{eq:spectral-dual}) numerically. This problem is solved via an interior-point method. The precise algorithm used is detailed in~\cite{Lawrence:2024hjm}. Here we note only that the algorithm is guaranteed to converge in polynomial time, due directly to the convex nature of the optimization problem.

\begin{figure}
	\centering
	\includegraphics[width=0.48\linewidth]{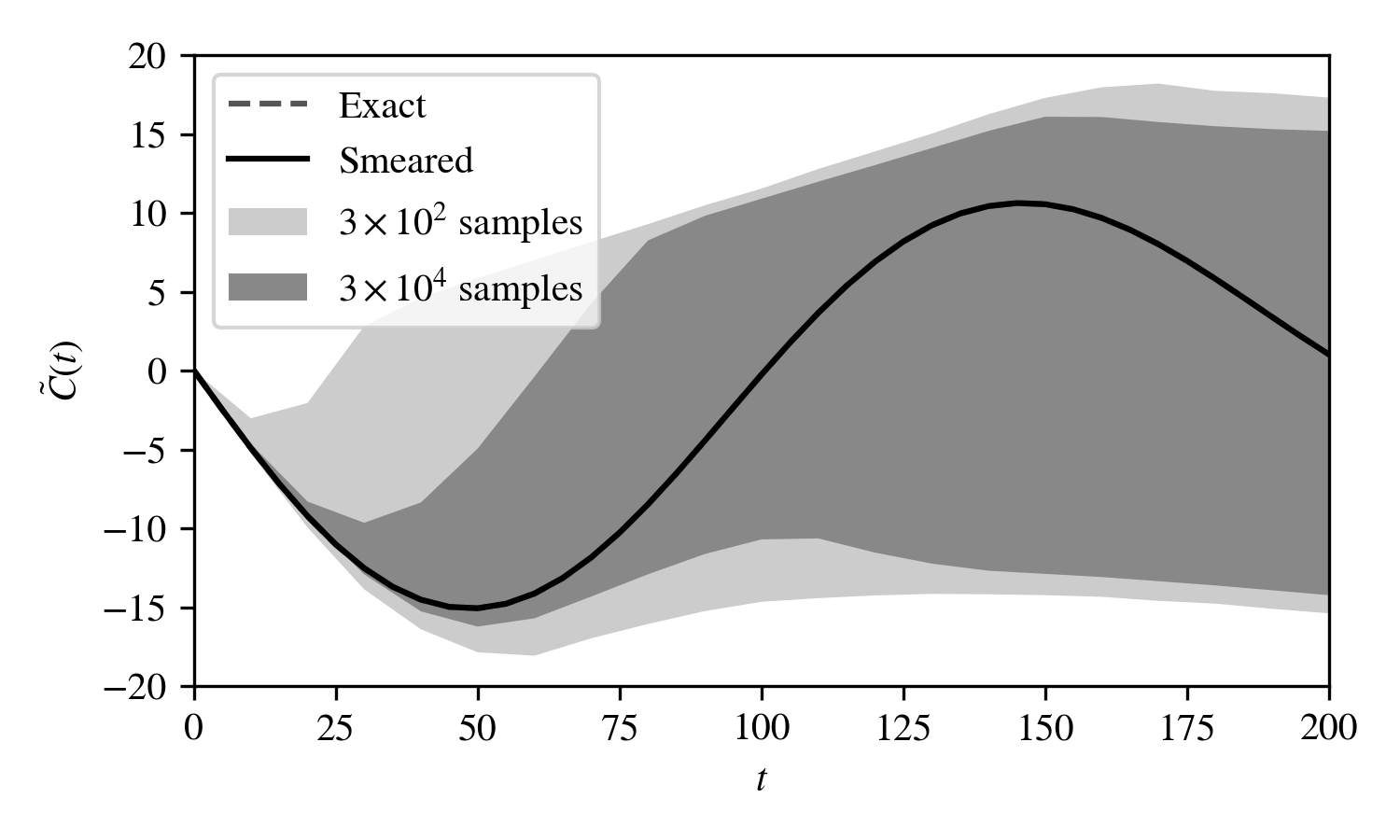}
	\hfill
	\includegraphics[width=0.48\linewidth]{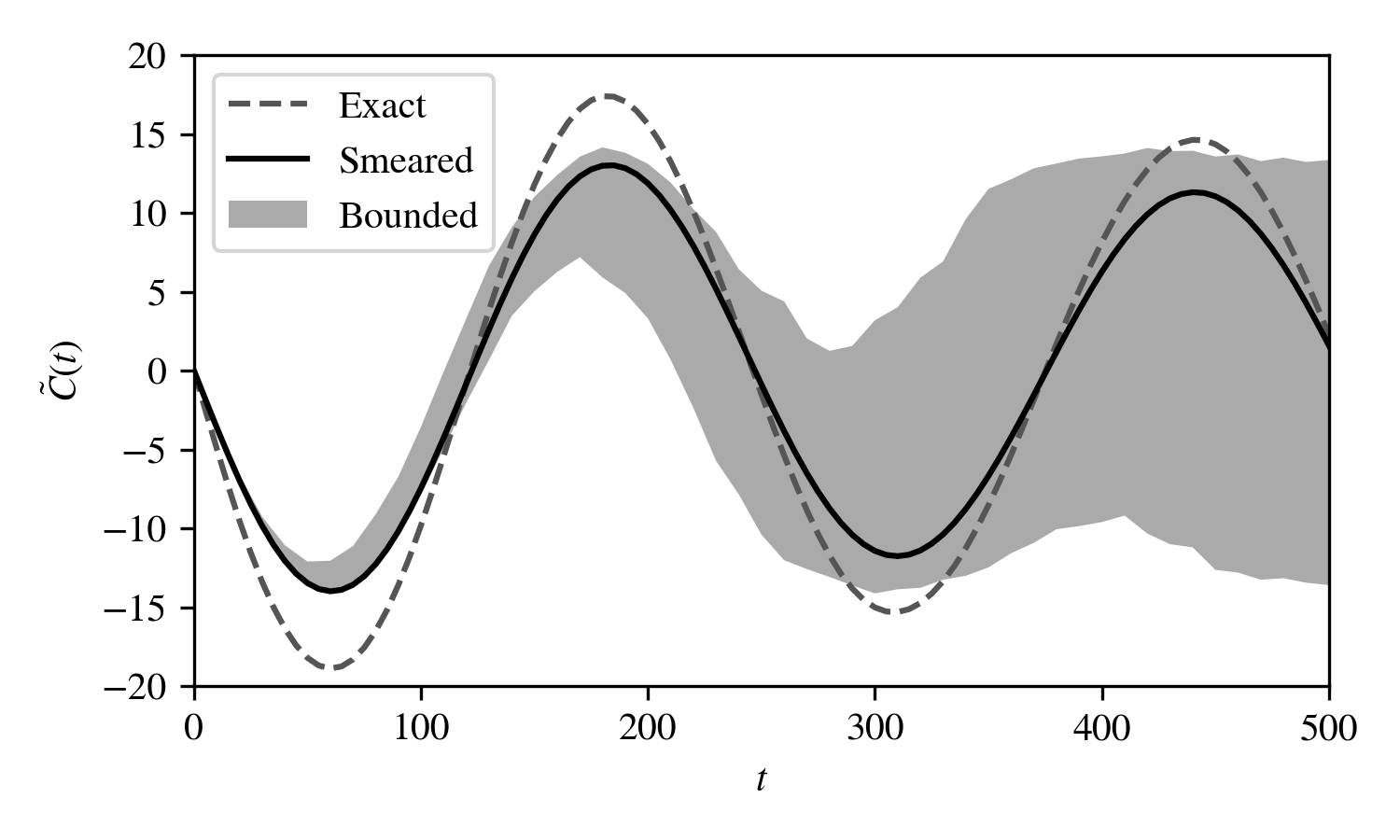}
	\caption{Extraction of the real-time correlator in the anharmonic oscillator ($\omega = 10^{-4}$ and $\lambda = 10^{-5}$) from noisy Euclidean data. At left is a reconstruction of the smeared correlator at an inverse temperature of $\beta = 30$. The two regions show how the bounds get somewhat tighter as the number of samples is increased. The smearing is defined by $\sigma = 1$. At right is a similar calculation but at much colder temperature $\beta=100$, and wider smearing $\sigma = 30$. Figure from~\cite{Lawrence:2024hjm}.\label{fig1}}
\end{figure}

To demonstrate (and in part, to check that the derivation of~(\ref{eq:spectral-dual}) is correct), consider the anharmonic oscillator. The Hamiltonian is
\begin{equation}
	\hat H_{\mathrm{osc}} = \frac 1 2 \hat p^2 + \frac{\omega^2}{2} \hat x^2 + \frac\lambda 4 \hat x^4
	\text,
\end{equation}
and we will take $\omega = 10^{-4}$ and $\lambda = 10^{-5}$. After obtaining $3 \times 10^4$ Monte Carlo samples, we solve the convex program (\ref{eq:spectral-dual}) at a large set of times, obtaining upper and lower bounds on the smeared real-time correlator at each time. Such bounds are shown in Figure~\ref{fig1}.

The left-hand panel of Figure~\ref{fig1} is from a lattice calculation performed at a relatively high temperature, with $\beta=30$. Here there is not enough lattice data to meaningfully constrain even one half-period of the real-time correlator. The situation is dramatically improved by a lowering the temperature to $\beta=30$ (shown in the right-hand panel), where a little more than a full period is visible from the bounds.

In both cases, the true smeared correlator (computed via direct diagonalization) is shown in a solid black line, and begins by closely following one of the two bounds. (Indeed, in the low-temperature case, the exact result closely follows the lower bound, and then after one half-period switches to nearly saturating the upper bound.) It is not obvious why this is the case, although it is weakly reminiscent of a common story in the conformal bootstrap, where interesting CFTs appear to lie very close to edges and corners of permitted regions in the space of scaling dimensions.

The performance of this method is not affected by the dimension of the lattice---only by the number of time-slices (and then only polynomially). Therefore with no additional effort beyond a more expensive Monte Carlo, we may extract real-time dynamics from a $2+1$-dimensional $\phi^4$ theory. This theory has a lattice action given by:
\begin{equation}\label{eq:scalar-action}
	S_{\mathrm{scalar}}(\phi) = \sum_{\langle r r'\rangle} \frac{\left(\phi_r - \phi_r'\right)^2}{2} + \sum_r \frac{m^2}{2} \left[\phi_r^2 + \frac{\lambda}{4} \phi_r^4\right]
	\text.
\end{equation}
We will take bare parameters $m^2 = 0$ and $\lambda = 10^{-2}$, producing an interacting scalar field with a mass gap $M \approx 10^{-1}$. The correlator we consider is $\langle \bar\phi(t) \bar\phi(0)\rangle$, where $\bar\phi$ is the spatial average of the field $\phi$:
\begin{equation}
 	\bar\phi(\tau) = \sum_x \phi_{x,\tau}
	\text.
\end{equation}
 
\begin{figure}
	\centering
	\includegraphics[width=3in]{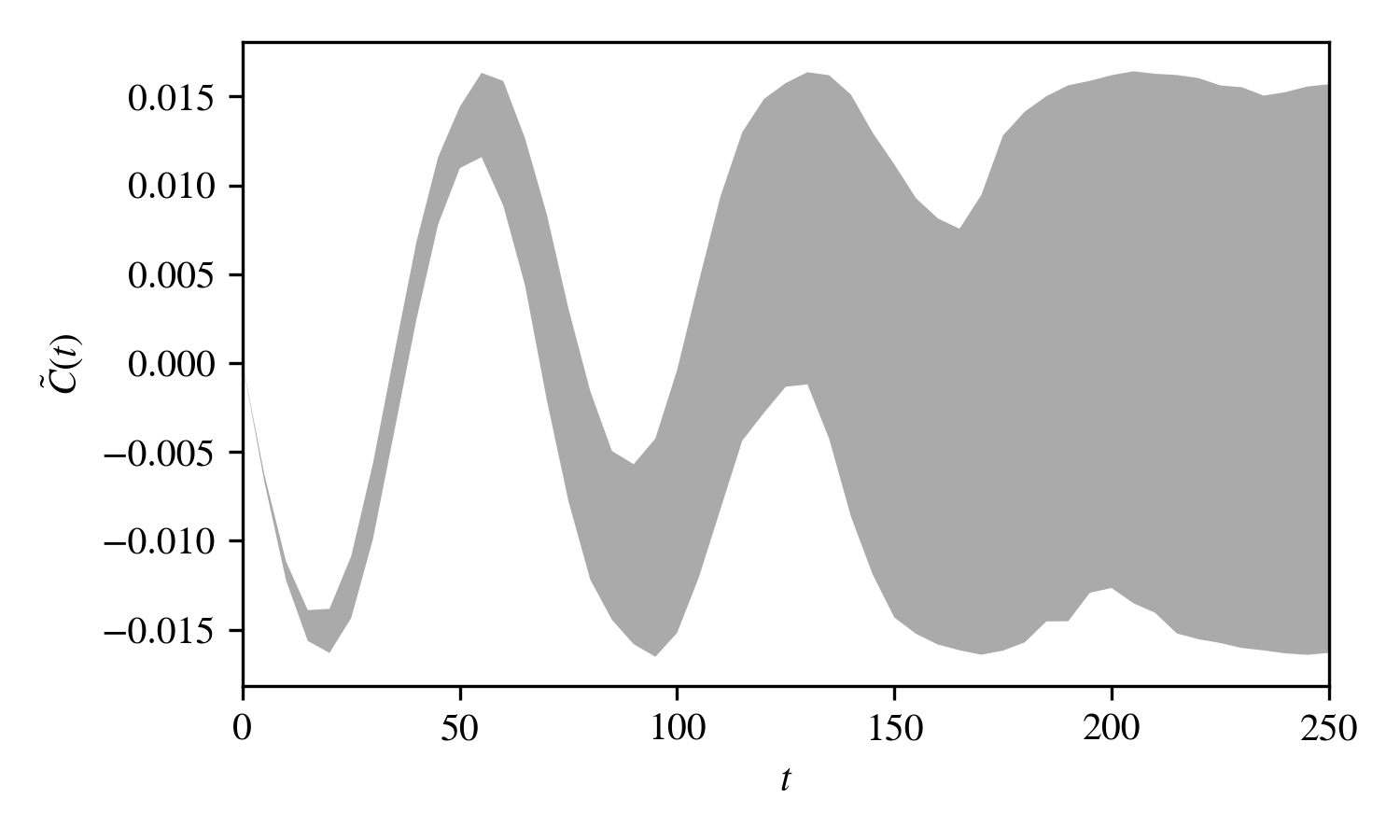}
	\caption{Extraction of a real-time correlator in $2+1$ $\phi^4$ theory, on a $16^2 \times 80$ lattice. The lowest-lying mass, in lattice units, is $M \approx 10^{-1}$. A total of $\sim 1.9 \times 10^5$ samples were used to determine the Euclidean correlator. Figure from~\cite{Lawrence:2024hjm}.\label{fig2}}
\end{figure}

Bounds on a smeared real-time correlator (with $\sigma = 30$) are shown in Figure~\ref{fig2}. The calculation is again done very near the ground state: $M\beta \approx 8$, where $M$ is the mass-gap on the lattice. With somewhat higher statistics, nontrivial bounds are available on the real-time correlator for nearly two full periods.

\section{Brief discussion}
The method presented here is quite restrictive. The only information used is the Euclidean correlator and the fact that $\rho(\omega) \ge 0$ at all frequencies (equivalent to reflection-positivity of the Euclidean lattice). The only quantities that can be extracted are linear functionals\footnote{Although, in principle this method may be extended to arbitrary convex functionals of $\rho$.} of the spectral density $\rho$; in other words, integrals of $\rho$. This excludes many physically natural statements about $\rho$, including transport coefficients, from being straightforwardly bounded. Ultimately this is not surprising, as in the absence of additional information about the theory, meaningful information regarding particle lifetimes or transport coefficients is in fact not present in the Euclidean correlator.

Of course, additional information \emph{is} available. To begin with, the lattice action is known, and therefore a (generally infinite) set of lattice Schwinger-Dyson relations is available. This allows the use of multiple correlation functions to more tightly constrain the spectral density. More powerful information is also at hand. In particular, the lattice is an approximation to a Lorentz-invariant field theory, and the Green's function therefore posess some approximately analytic structure in the form of poles and branch points that convey physical information (regarding, depending on the correlator chosen, lifetimes, S-matrix elements, or quasinormal modes).

In exchange for disregarding all of this information, we have gained an efficient algorithm which provides entirely model-independent bounds on arbitrary integrals of the spectral density.

To what extent is any of this useful for lattice QCD? In practice, here lies a minor conundrum. We have assumed reflection-positivity throughout. If a correlator that does not obey $\rho(\omega) \ge 0$ is fed into the method described above, one obtains incorrectly tight bounds\footnote{In practice, the method is often able to prove that no positive spectral density function could have produced the given correlator. Numerically this manifests as a lower bound which exceeds the upper bound.}. In other words, the method is worth less, and perhaps worthless, unless the correlator comes from an exactly reflection-positive lattice. Application to lattice QCD will either be restricted to the use of such correlators, or require that the assumption of strict positivity be partially lifted.

\acknowledgments
	The work described in this talk was preceded and informed by many insightful conversations with Tanmoy Bhattacharya, Tom Cohen, Tom DeGrand, Brian McPeak, Duff Neill, and Paul Romatschke.

This work was supported by a Richard P.~Feynman fellowship from the LANL LDRD program. Los Alamos National Laboratory is operated by Triad National Security, LLC, for the National Nuclear Security Administration of U.S. Department of Energy (Contract Nr. 89233218CNA000001). Report number: LA-UR-25-20001.

\bibliographystyle{JHEP}
\bibliography{rt}

\end{document}